\documentclass[sigconf,natbib=true]{acmart}
\AtBeginDocument{%
  \providecommand\BibTeX{{%
    \normalfont B\kern-0.5em{\scshape i\kern-0.25em b}\kern-0.8em\TeX}}}

\setcopyright{acmcopyright}
\copyrightyear{2023}
\acmYear{2023}
\acmDOI{xx.xxxx/xxxxxxx.xxxxxxx}

\acmConference[WSDM '24]{WSDM '24: The 16th ACM International Conference on Web Search and Data Mining}{February 27-March 3, 2024}{Mexico}
\acmBooktitle{WSDM '24: The 17th ACM International Conference on Web Search and Data Mining, February 27-March 3, 2024, Mexico}
\acmPrice{15.00}
\acmISBN{978-1-4503-XXXX-X/18/06}

\usepackage[utf8]{inputenc}
\usepackage{amsmath}
\usepackage{amsfonts}
\usepackage{multirow}

\usepackage{multirow}
\usepackage{graphics}
\usepackage{balance}
\usepackage{dsfont}
\usepackage{caption}
\usepackage{subcaption}
\usepackage{algorithm}
\usepackage{algorithmic}

\newcommand{\ie}{\textit{i.e.}}
\newcommand{\eg}{\textit{e.g.}}

\newcommand{\todo}[1]{\{\textcolor{blue}{\textbf{TODO}}\}}

\title{Session-level Normalization and Click-through Data Enhancement for Session-based Evaluation}

\author{Haonan Chen}
\affiliation{%
  \institution{Gaoling School of Artificial Intelligence, Renmin University of China}
  \city{Beijing}
  \country{China}}
\email{hnchen@ruc.edu.cn}

\author{Zhicheng Dou}
\affiliation{%
  \institution{Gaoling School of Artificial Intelligence, Renmin University of China}
  \city{Beijing}
  \country{China}}
\email{dou@ruc.edu.cn}

\author{Jiaxin Mao}
\affiliation{%
  \institution{Gaoling School of Artificial Intelligence, Renmin University of China}
  \city{Beijing}
  \country{China}}
\email{maojiaxin@ruc.edu.cn}

\begin{document}

\begin{abstract}

Since a user usually has to issue a sequence of queries and examine multiple documents to resolve a complex information need in a search session, researchers have paid much attention to evaluating search systems at the session level rather than the single-query level. Most existing session-level metrics evaluate each query separately and then aggregate the query-level scores using a session-level weighting function. The assumptions behind these metrics are that all queries in the session should be involved, and their orders are fixed. However, if a search system could make the user satisfied with her first few queries, she may not need any subsequent queries. Besides, in most real-world search scenarios, due to a lack of explicit feedback from real users, we can only leverage some implicit feedback, such as users' clicks, as relevance labels for offline evaluation. Such implicit feedback might be different from the real relevance in a search session as some documents may be omitted in the previous query but identified in the later reformulations. To address the above issues, we make two assumptions about session-based evaluation, which explicitly describe an ideal session-search system and how to enhance click-through data in computing session-level evaluation metrics. Based on our assumptions, we design a session-level metric called Normalized U-Measure (NUM). NUM evaluates a session as a whole and utilizes an ideal session to normalize the result of the actual session. Besides, it infers session-level relevance labels based on implicit feedback. Experiments on two public datasets demonstrate the effectiveness of NUM by comparing it with existing session-based metrics in terms of correlation with user satisfaction and intuitiveness. We also conduct ablation studies to explore whether these assumptions hold.

\end{abstract}

%\begin{CCSXML}
%<ccs2012>
%   <concept>
%       <concept_id>10002951.10003317.10003359</concept_id>
%       <concept_desc>Information systems~Evaluation of retrieval results</concept_desc>
%       <concept_significance>500</concept_significance>
%       </concept>
%\end{CCSXML}

%\ccsdesc[500]{Information systems~Evaluation of retrieval results}

%\keywords{Session-level Normalization, Click-through Data Enhancement, Session Search, Evaluation Metrics}

\maketitle

\section{Introduction}\label{sec:intro}

With the development of search engines, researchers increasingly focus on building better evaluation methods. In the early years, the Cranfield paradigm~\cite{cranfield} was the dominant approach in evaluating the search results of a single query. 
However, when a user is trying to complete a complex search task, she may issue multiple queries and browse a series of documents to obtain sufficient information in a \emph{search session}~\cite{coca, hqcn, ricr}. 
Many works have emerged to design session-based evaluation metrics.
Some of them have already been used in some evaluation tasks, \eg, Session-based DCG~\cite{sdcg} in the TREC Session Track~\cite{TREC2010} and Recency-aware Session-based Metric~\cite{rsdcg} in the recent NTCIR Session Search (SS) Task~\cite{ntcir}. 
However, there are still some remaining challenges for session-based evaluation. In this work, we identify two major challenges of session-based evaluation as some existing session-based metrics are based on oversimplified or problematic assumptions.  

\textbf{The first challenge} is that most existing session-level metrics (\eg, the metrics used in these tasks) evaluate each query in a session based on an existing query-level metric, and then aggregate those query-level scores with some session-level discount factors or weighting schema to evaluate systems at the session level.
For example, sDCG~\cite{sdcg} is based on the cascade hypothesis, which gives lower-ranked search results and later-issued queries smaller weights, and RSMs~\cite{rsdcg} gives larger weights to the recently issued queries. 
While previous studies show that these aggregation metrics correlate well with users' session-level satisfaction feedback~\cite{rsdcg}, these metrics all implicitly assume that a user's query sequence in a search session will not be altered by the systems. 
However, it would be expected that if a user is satisfied with the information retrieved by the present and past queries, she may not need any subsequent queries. (Following~\cite{jones2008, Wang2013}, we assume that in the same session, the queries that may represent different sub-topics serve the same primary information need.)
To put it another way, we presume that an ideal search system would return all relevant documents in a session before all irrelevant documents, so the user can spend the least effort in completing the search task. 
For example, a user issues a query ``Java'' and clicks ``What is Java Language''. After a minute, she issues another query ``Java Project'', and clicks ``Java Projects for Beginners''. 
We assume the user would prefer a system if it could predict that she is seeking the second clicked document (\eg, using a personalized search model) and place it at the beginning of this session.
However, none of the existing session-level evaluation metrics can fully take this reduction in effort into consideration and give a maximum score to such an ideal search system. 
Note that it is true that a user may learn something in the session which triggers her to seek new pieces of information. 
However, we believe that an ideal system can predict the change of her interest, lead her to discover all information needs as soon as possible, and rank all the documents that serve her needs high in the session.

\textbf{Another challenge} that may hinder the computation of session-level evaluation metrics is that in real-world search scenarios, search engines can only record implicit user feedback to the documents, \ie, click-through data. 
Therefore, most metrics have to assume that the clicked documents are ``relevant'' to the query in offline evaluation.
However, users may skip some relevant documents during the search because of position bias. 
This problem is more common in a long search session as Price et al.~\cite{DBLP:conf/cikm/PriceNDV07} found out that users may omit documents in previous queries yet recognize and click them in later reformulations, \ie, the session's subsequent queries. 
For example, a user issues a query ``MacBook'' then clicks ``MacBook on Amazon''. A minute before, the search system had already ranked this document among the top ten results when she searched ``Apple'', but she omitted it. 
We assume that when evaluating this system, we should also mark "MacBook on Amazon" relevant to "Apple" in order to reward this system for successfully predicting the user's search intent and saving her efforts on issuing another query. 
Thus, we need to consider the ``relevance'' of a document at the session level but not at the level of each separate query.
However, existing metrics do not account for these omitted documents and simply assume the unclicked documents are irrelevant (when only implicit feedback is available), because these metrics only take clicks as per-query relevance judgments, rather than considering clicks in the entire session.

%\textbf{(2)} There may be duplicate documents in a multi-query search session, so the same document could be counted more than once during evaluation.
%For example, a user's first query is ``Homemade Cookie'' and she clicks ``How to make cookie at home?''. After a minute, she issues another query ``What desert can be made at home?'' and clicks the same document again. 
%In this case, she is attempting to explore different kinds of desserts to make at home, and she believes that this document is a new type of cookie or a new way to prepare a cookie because it is ranked among the top results again, despite the fact that it contains no new information.
%As a result, we assume that because she has gotten all the information in this document, it should be marked irrelevant to her subsequent queries. Then the evaluation results can reward the search system that can return novel and more diversified results to the user.
%Among the existing works, some do not give any special treatment to these duplicate relevant documents~\cite{nsdcg, umeasure} and some choose to discount these documents~\cite{DBLP:conf/ictir/YangL09}. 
%However, none of these works have empirically investigated how it may affect the session-level evaluation. 
%In Section~\ref{subsec:abla}, we demonstrate that this assumption is actually \textbf{invalid}. (A possible reason is that the user may need to re-find information that she had already found or skimmed through earlier in the session.)

To tackle these challenges, in this paper, we make two \textbf{assumptions} about a user's search behaviors in a search session: 

\textbf{\underline{Assumption 1}}: An ideal search system should rank all the documents that the user requires, \ie, all the relevant documents, before all the irrelevant documents in the entire session. By doing so, it can save the user's effort because she may not even have to reformulate her queries. In other words, we need to remove the boundary of queries in the session-level evaluation.

\textbf{\underline{Assumption 2}}: 
A document that is clicked in a subsequent query but omitted in a preceding query in the same session is relevant to that preceding query. Consequently, the first occurrence of this document is assumed to be relevant to the information need of the current session.

% \textbf{\underline{Assumption 3}}: 
%If the user has already clicked a document in a preceding query, this document is not relevant to her subsequent queries because of redundancy.

\noindent We will refer to Assumption 1 as A1 and Assumption2 as A2.

Most offline evaluation in the industry can only use implicit user feedback (click-through data) to infer relevance labels because it is costly to get human relevance labels.
Besides, most human-labeled relevance is query-level relevance, not the session-level relevance that is preferred when evaluating session search systems. 
Thus, our assumptions in this paper are based on actual click-through data (implicit feedback) and user behaviors rather than manual labels. 
Under this condition, we describe an ideal session search system (A1) and try to mine session-level relevance labels from click-through data (A2). 

The common idea implied in these assumptions is that we need to reduce the impact of query boundaries in session search evaluation. 
%Document relevance derived from clicks in a session should go across the original queries issued by the user. 
Based on these assumptions, we design a session-level metric called \textbf{Normalized session level U-Measure (NUM)} based upon the original U-Measure~\cite{umeasure}.
This metric evaluates a session as a single ``virtual query''  and employs an ideal session to normalize the evaluation result according to {A1}. 
NUM also converts the click-through feedback into session-level relevance labels based on {A2}. 
%We do not calculate per-query scores in NUM. 
%Instead, it will directly yield the session-level discounted cumulative gain.
Experiments on two public datasets (TianGong-SS-FSD~\cite{rsdcg} and NTCIR-16 Session Search Task~\cite{ntcir}) show that NUM is an intuitive session-level metric that correlates well with user satisfaction. 
Furthermore, ablation studies confirm that 
%NUM is more intuitive and correlates better with user satisfaction with {A1} and {A2}, which demonstrates that 
{A1} and {A2} are \textbf{valid}.
%, but its performance decreases with {A3} (\ie our ablation studies demonstrate that {A1} and {A2} are \textbf{valid}, whereas A3 is \textbf{invalid}).

To summarize, the contributions of the paper are as follows:

(1) We make two assumptions about what we should do in session-based evaluation. With these assumptions, we describe what an ideal session should be and discuss how to use the click-through data to derive session-level relevance labels. 

(2) We design a session-level metric called NUM based on U-measure~\cite{umeasure}. It treats a session as a virtual query and uses the evaluation score of the ideal session for normalization (\textbf{session-level normalization}).

(3) We show that NUM correlates well with user satisfaction. %than the state-of-the-art metrics. 
Further studies also demonstrate its intuitiveness. In addition, the ablation studies explore the reasonability of our two assumptions.
%
%\end{itemize}
\begin{figure*}[tbp!]
    \centering
    \includegraphics[width=0.8\textwidth]{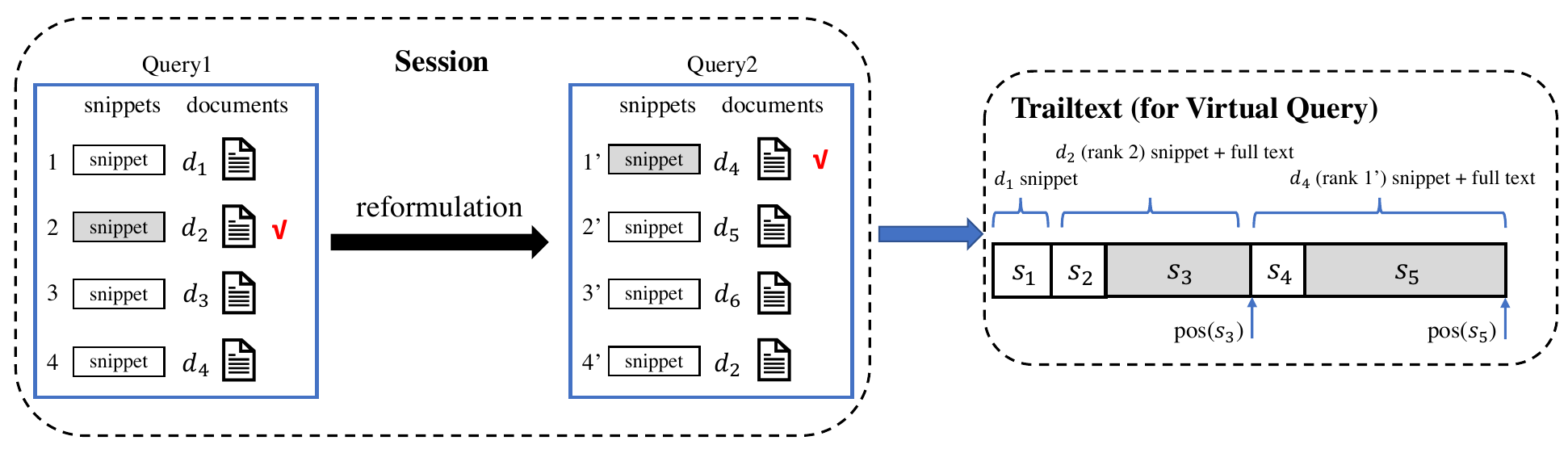}
    \caption{The illustration of how U-measure constructs trailtext from a two-query session. The results clicked by the user are marked with \textcolor{red}{red} checkmarks and the results marked as relevant are filled with color \textcolor{gray}{gray}. The right part is the constructed trailtext, where ``$s_i$'' is the $i$-th string of it.}
    \label{fig:um-tt}
\end{figure*}
\section{Related Work}\label{sec:RW}

\subsection{Session-based Evaluation Metrics}

Because in real-world search scenarios, users usually issue a sequence of queries to complete a complex search task, researchers have gone beyond the traditional query-level metrics (\eg, MAP, nDCG, and MRR) and designed some session-level metrics.

Session-based DCG (sDCG)~\cite{sdcg} is a multi-query metric based on Discounted Cumulative Gain (DCG)~\cite{dcg}. Based on the cascade hypothesis, sDCG discounts the weights of the lower-ranked results and later-issued queries. 
Similarly, Lipani et al.~\cite{srbp} proposed a session-based version of Rank-Biased Precision (RBP)~\cite{rbp}, which added a new parameter over RBP to balance between query reformulation behaviors or continuing to examine documents. 
Yang and Lad~\cite{DBLP:conf/ictir/YangL09} proposed a utility-based evaluation framework. They evaluated the Expected Utility of the search system over all possible interaction patterns.
Van Dijk et al.~\cite{conf/ecir/DijkFFK19} leverages a Markovian Chain of users' behaviors in a session to substitute the fixed discount of documents.
Moffat et al.~\cite{DBLP:journals/ipm/WicaksonoM21}, extended the C/W/L framework to session-based effectiveness evaluation.

Liu et al.~\cite{DBLP:conf/sigir/LiuLMLM18, DBLP:conf/kdd/LiuMLZM19} showed how the recency effect can affect users' session-level satisfaction with user studies. They proposed that the later-issued queries should receive higher weights, which is in contrast to the cascade hypothesis. 
Based on these findings, Zhang et al.~\cite{rsdcg} proposed Recency-aware Session-based Metrics (RSMs), which incorporate the recency effect into session-based evaluation. 
%They further constructed a field study dataset called TianGong-SS-FSD and meta-evaluated two instantiations of RSMs (RS-DCG and RS-RBP) and various session-based metrics, including sDCG and sRBP. 
Their experimental results showed that RSMs have the strongest correlations with user satisfaction among these metrics and achieve state-of-the-art performance in estimating user satisfaction. 

Although various session-based evaluation metrics have been proposed, and some perform quite well in estimating user satisfaction and/or measuring system effectiveness, most of these session-based metrics evaluate each query separately and aggregate the evaluation scores using some discount factors or weighting schema. 
Therefore, they all implicitly assume that all queries in the session should be considered in the session-based evaluation and the orders of the queries are definite. 
However, as we stated in \textbf{A1}, if a search system could make the user satisfied with the first few queries, she may not need any subsequent queries. Few existing session-based metrics have taken this into consideration.

Besides, most existing metrics are based on the Cranfield/test collection approach (test collections with explicit relevance judgments). 
However, as stated in Section~\ref{sec:intro}, most offline scenarios in the industry can only use implicit user feedback (click-through data). Therefore, these metrics would have to directly treat click-based labels as relevance labels.
Our work attempts to bridge this gap by enhancing click-through data based on \textbf{A2}.
By this, we aim to infer session-level relevance labels with implicit feedback.

\subsection{U-measure}\label{subsec:RW-um}

In this part, we will briefly review U-measure~\cite{umeasure}, based on which we design a session-level metric NUM. U-measure is a framework for evaluating information access that can be used to evaluate various IR tasks. 

Figure~\ref{fig:um-tt} is an example in its original paper of how U-measure constructs a trailtext from a two-query session. 
When evaluating a session, U-measure treats the session as a single ``virtual query'' (as a whole) by building a \textbf{trailtext} for this session. 
A trailtext $tt$ is made up of $n$ strings concatenated together: $tt = s_1 s_2 \cdots s_n$. 
Each string $s_i$ could be a snippet or the entire content. Sakai and Dou presume that the trailtext represents what the user reads in the exact order during an information search. 
Besides, they assume that a user reads $F$\% (20\% in~\cite{umeasure}) of the content of a document, \ie, only $F$\% of a document's length is counted in the trailtext.
They define $pos(s_i) = \sum_{j=1}^i |s_j|$ as the offset position of $s_i$.
Specifically, trailtexts are derived from session data that follows the algorithm in~\cite{umeasure} (Figure 5).
The general computation of U-measure is:
\begin{equation}
    {\rm U} = \sum^{|tt|}_{pos=1} {\rm gain}(pos){\rm D}(pos), \label{Eq:umeasure-1}
\end{equation}
where ${\rm D}(pos)$ is the position-based decay function and ${\rm gain}(pos)$ is the corresponding gain. Specifically, if $s_i$ is not relevant, ${\rm gain}(pos(s_i))$ = 0, and if $s_i$ is regarded $l$-relevant, ${\rm gain}(pos(s_i)) = gv_l$, where $gv_l$ is a gain value for relevance level $l$. ${\rm D}(pos)$ is a linear decay function:
\begin{equation}
    {\rm D}(pos) = {\rm max}(0, 1-\frac{pos}{L}),  \label{Eq:umeasure-2}
\end{equation}
where $L$ denotes the maximum amount of text the user may read in a single session and represents the point at which all information units are considered worthless. 

We choose U-measure as our metric's backbone because: (1) Rather than aggregating the evaluation results of all queries in a session, which may contain queries that the user does not need, U-measure evaluates a session as a single ``virtual query'' by building a trailtext. (2) U-measure takes the length of a document into consideration and has the diminishing return property, making it more realistic than rank-based metrics. 

Note that the original paper of U-measure~\cite{umeasure} primarily introduces it as a general evaluation framework for a variety of IR tasks, rather than specifically for session search. 
Therefore, many improvements are required to bring it to session-level evaluation (based on our two assumptions). 
Experiments conducted in Section~\ref{subsec:abla} demonstrate the effectiveness and necessity of these modifications. 

\section{The Proposed Methods}\label{sec:Method}
%First of all, to address the challenges we discussed in Section~\ref{sec:intro}, we make the following three assumptions for session-based evaluation:

In this section, we will introduce the proposed metric \textbf{Normalized U-Measure (NUM)}. When evaluating a session, instead of aggregating the scores of each query in the session, it evaluates the session as a ``virtual query'' by building a comprehensive trailtext for the whole session. 
Besides, it uses the ideal session to normalize the evaluation result in accordance with \textbf{A1}. 
Based on \textbf{A2}, it uses click-through data to infer session-level relevance labels.

\begin{figure*}[tbp!]
    \centering
    \includegraphics[width=0.85\textwidth]{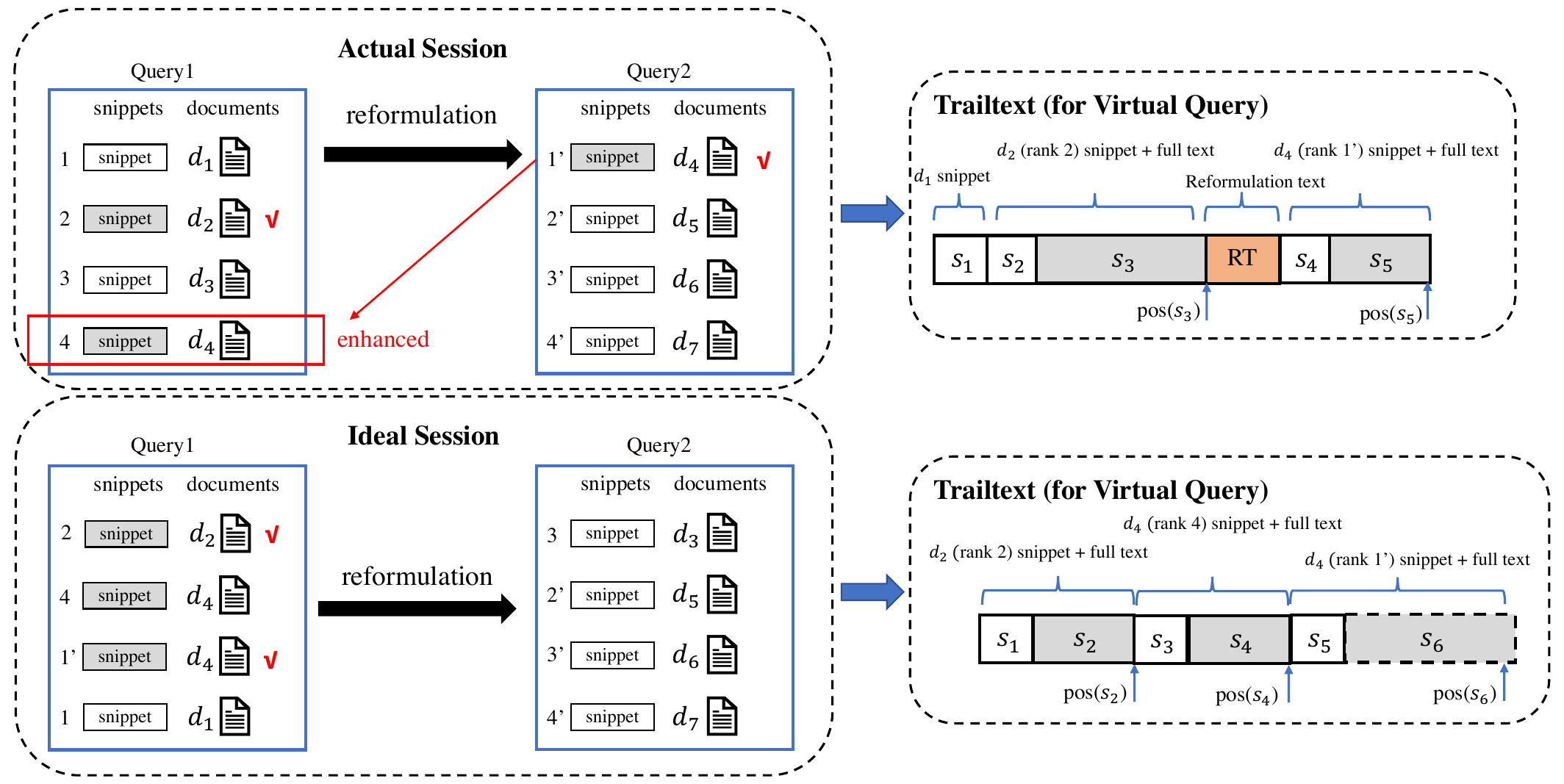}
    \caption{The illustration of NUM. The upper part is the actual session, and the lower part is the ideal session. The results clicked by the user are marked with \textcolor{red}{red} checkmarks, and the results marked as relevant are filled with color \textcolor{gray}{gray}. 
    Rank 4 of the first query is marked relevant even though it is not clicked here because it is clicked in the subsequent query. 
    %Rank 4' of the second query is marked irrelevant even though it is clicked here because it has been clicked in the preceding query. 
    %Note that we put $d_4$ before $d_2$ in the ideal session because $d_4$'s text length is shorter.
    We treat a session as a virtual query, based on which we build a trailtext to enable session-level evaluation.
    We construct the trailtext based on actual user clicks for the actual session and based on the enhanced session-level labels for the ideal session.}
    \label{fig:num}
\end{figure*}

\subsection{Normalizing U-Measure}

\subsubsection{Ideal Session} \label{subsubsec:ideal_session}
As we stated in {A1}, we believe that an ideal search system should rank all relevant documents before all irrelevant documents so that the user may not even need to reformulate her queries, saving her time and effort.
As illustrated in Figure~\ref{fig:num}, 
%we try to evaluate the whole session as a virtual query by building a trailtext.
we place all the relevant documents before all irrelevant documents to build the ideal trailtext because it allows the user to get all of the information she needs without reformulation.
By doing so, we manage to explicitly tell the search system what an optimal session-level ranking is and lead it to limit the number of reformulations, which we believe will make the user more satisfied. 
Note that the relevance label we refer to here is the session-level labels derived from click-through data based on A2.

We notice that there could be different re-ranking strategies of the documents with the same relevance scores in the ideal session.
If the user scans the ranked documents from top to bottom, she will learn by reading the preceding document that subsequent documents are relevant.
Thus, we maintain the same order of clicks in the trailtext of the ideal session as the actual session.
%To have a more straightforward view of this, let us take an example. Let us suppose that there are two relevant documents: $d_4 (l_4 = 1000), d_2 (l_2 = 2000)$, and $F=20, L=2000$. 
%Then there are two different re-ranking strategies: $[d_4, d_2]$ or $[d_2, d_4]$. 
%Based on Eq.~(\ref{Eq:umeasure-1}) and Eq.~(\ref{Eq:umeasure-2}), we will have: $U_{[d_4, d_2]} = (1-200/2000) * 1 + (1-600/2000) * 1 = 1.6$ and $U_{[d_2, d_4]} = (1-400/2000) * 1 + (1-600/2000) * 1 = 1.5$. 
%(Following~\cite{umeasure}, we assume 20\% of a document's context is read.)
%We can find that the order of the relevant documents with different lengths can affect the calculation of U-measure of the ideal session.
%Based on their text lengths, we will re-rank the documents with the same relevance level in ascending order. 
%This operation will maximize the evaluation score given by Eq.~(\ref{Eq:umeasure-1}) and Eq.~(\ref{Eq:umeasure-2}), thus restrict the result of NUM in [0,1]. 

\subsubsection{Session-level Normalization} \label{subsubsec:normalize}
The value of NUM represents the proximity of an actual session to the ideal session. 
Note that when building trailtext for the actual session, based on~\cite{JoachimsGPHRG07, umeasure}, we assume that the user reads all the documents she clicks and all the snippets above the lowest clicked document in the session.
And for the ideal session, we simply presume the user reads the documents that are marked session-level relevant (inferred by click-through data and A2) since all of them are presented at the beginning of the session and they all support the user's main information need.
The session-level normalization technique is defined as follows: 
\begin{equation}
    {\rm NUM} = \frac{{\rm U}(S_{\rm actual})}{{\rm U}(S_{\rm ideal})},\label{Eq:num}
\end{equation}
as illustrated in Eq~(\ref{Eq:umeasure-1}) and Eq~(\ref{Eq:umeasure-2}), ${\rm U}(S)$ can be computed as follows:
\begin{equation}
    {\text{U}}(S) = \sum_{pos=1}^{|{tt}|} {\rm gain}(pos) {\rm max}(1-\frac{pos}{L}), \label{Eq:full-um}
\end{equation}
where $L$ is the longest Maximal Trailtext Length (MTL) across all conceivable search sessions. 
A session's MTL is the sum of the text lengths of (1) all snippets above the last click of each query, (2) all documents clicked by the user in that session, and (3) all reformulation texts (illustrated in Section~\ref{subsubsec:rt}). 
As a result, $L$ reflects the most text the user has to read in a single session.

\subsubsection{Reformulation Text} \label{subsubsec:rt}

As shown in the upper right part of Figure~\ref{fig:num}, we add an empty text named \textbf{reformulation text} in the trailtext between two queries.
We believe it can penalize query reformulation behaviors to an adjustable degree, which the original U-measure does not account for. The intuition here is that U-measure uses text length to simulate reading time, thus we use empty texts to represent the user's reformulation time. 

We determine the length of the reformulation text by exploring a Chinese field study dataset TianGong-SS-FSD~\cite{rsdcg} to find out how much time users spend between the exit of a query and the issuance of the following query. 
%The detail of this dataset is presented in Section~\ref{subsec:datasets}. 
We compute the query reformulation time for each query. 
Besides, we eliminate the queries with negative reformulation time (due to multi-tabbing or logging errors) and 4\% of queries with the largest reformulation time values (extremely long times for unknown reasons).
More details are in Section~\ref{subsec:satisfaction}.

\subsection{Click-through Data Enhancement} \label{subsec:click-enhancement}

We use the click-through data to infer session-level relevance labels based on \textbf{A2}.
%we adopt a type of enhancement to explore this assumption:
%Specifically, we enhance preceding queries' labels based on subsequent queries' clicks.
We believe the same document should be marked relevant to this session if it was clicked in a subsequent query but skipped in a preceding one.
As illustrated in the left part of Figure~\ref{fig:num}, the document \textbf{$d_4$} at rank 4 is marked relevant because it is clicked in the subsequent query. 
%Note that we construct trailtext for the real session from actual user clicks (illustrated in the upper right part of Figure~\ref{fig:num}) and from relevance judgment for the ideal session (illustrated in the lower right part of Figure~\ref{fig:num}). 
%It's because the trailtext is what the user actually sees, but we can't get the user's view for the ideal session because the rankings of the documents have changed. 
%As a result, we can only presume that the user will examine all of the documents that we consider relevant to the session (the documents the user has clicked in the actual session and those modified as relevant).

%(2) \textbf{Enhancing the subsequent queries' labels based on the preceding queries' clicks.} According to \textbf{A3}, if the same document is clicked in a preceding query and also clicked in a subsequent query, it should be marked irrelevant to the preceding one.
%As illustrated in the middle part of Figure~\ref{fig:num}, the document \textbf{$d_2$} at rank 4' is marked irrelevant even though it is clicked here because it has already been clicked in the preceding query.
%Although we include it in the trailtext of the actual session (since the user did click it), the gain value of it is still 0 because it has been modified as irrelevant to this session. (In Section~\ref{subsec:abla}, we actually find that A3 is \textbf{invalid}.)

Note that there will be duplicate documents that are considered relevant in the ideal session after the enhancement, \eg, $d_4$ at rank 4 and rank 1'.
Among the existing works, some do not give any special treatment to duplicate relevant documents~\cite{nsdcg, umeasure} and some choose to discount these documents~\cite{DBLP:conf/ictir/YangL09}. 
%However, none of these works have empirically investigated how it may affect evaluation. 
We identify three possible choices of dealing with $d_4$: 
(1) We include it in the trailtext because we believe that both rank 4 and rank 1' are informative. 
Besides, users may hope to find their clicked documents remain in the top results, which can facilitate the re-finding behaviors and the trust of search systems~\cite{nsdcg}; 
(2) We include it in the trailtext but give a discount to its relevance score;
(3) We exclude it in the trailtext because it is redundant. 
%We believe that both choices are reasonable, and we include $s_4$ in our instantiation of NUM.
We will include $s_6$ in our instantiation of NUM (choice (1)) and study these three approaches in Section~\ref{subsec:duplicate}.

%Note that we include rank 1' ($s_4$) in the trailtext of the ideal session. In other words, we do not discard it based on \textbf{A3} because rank 4 has been enhanced to be relevant by rank 1' based on \textbf{A2}. We believe rank 1' is still informative because it has actually been clicked by the user in the actual session.

\section{Experimental Setup}

\subsection{Datasets} \label{subsec:datasets}

We conduct our experiments on two public datasets: TianGong-SS-FSD~\cite{rsdcg} and NTCIR-16 Session Search (SS) Task~\cite{ntcir}. For simplicity, we denote the two datasets as $FSD$ and $NST$ in what follows\footnote{We do not use TREC 2014 Session Track because its sessions are relatively old and we want to study user behaviors on modern search engines.}. 

\subsubsection{$FSD$} 

$FSD$ is a dataset collected from field studies.
It records the users' session-level satisfaction rating, which we believe is a good standard for evaluating the metrics because it reflects the user's actual feelings about the search system. We explore the correlation of metrics (NUM and the baseline metrics) with user satisfaction ratings on this dataset.
Following~\cite{rsdcg}, we filter out the sessions containing more than one SERP.
Besides, to facilitate NUM usage, we filtered out the sessions that do not contain any clicks and only have one query. 
We believe this kind of session can be considered \textbf{good abandonment} because in each of these sessions, the user's information need is resolved only by the results pages (\eg, the information of snippets), without having to click on a result or do any query reformulations~\cite{good_abandonment}.

\begin{table}[t!]
    \centering
    
    \caption{The statistics of the pre-processed datasets.}
    \begin{tabular}{l@{}rr}
    \toprule
         & \textbf{TianGong-SS-FSD} & \textbf{NTCIR-16 SS} \\
    \midrule
        \# Sessions & 994 & 2,000  \\
        \# Queries  & 3,411 & 6,420  \\
        Avg. \# Query per Session & 3.43 & 3.21 \\
        \# Results per Query & 10 & 10  \\
        Avg. \# Clicks per Session & 3.49 & 3.08 \\
    \bottomrule
    \end{tabular}
    \label{tab:sta}
    \vspace{-5px}
\end{table}

\subsubsection{$NST$} 

$NST$ is collected from a Chinese search engine. Its full data has 147,154 sessions, and there are 2,000 of them that have human-labeled relevance. 
We test the intuitiveness of the metrics on these 2,000 sessions because we may use these manual labels in future work and the scale of 2,000 sessions are large enough to draw conclusions (already larger than the dataset used in many works that are based on Cranfield/test collection~\cite{rsdcg, srbp, sdcg}). 
We use the remaining 145,154 sessions to estimate $L$ and the length of the reformulation text when testing intuitiveness on $NST$.
These sessions do not contain any good abandonment.

The statistics of the pre-processed datasets are shown in Table~\ref{tab:sta}. 

\subsection{Meta-evaluation Approaches}

We utilize two meta-evaluation approaches to evaluate and compare NUM with existing session-based metrics:

\subsubsection{Correlation with User Satisfaction.}

Since user satisfaction in information retrieval can be defined as the fulfillment of a specific objective~\cite{DBLP:journals/ftir/Kelly09} and it assesses users' actual feelings about a system, it can be considered as the ground truth to evaluate the evaluation metrics~\cite{DBLP:conf/sigir/Al-MaskariSC07, DBLP:conf/sigir/HuffmanH07}. 
With the session-level user satisfaction feedback in $FSD$, we can compare the performance of different session-based metrics by computing the correlation between them and user satisfaction on this dataset. 

%\subsubsection{Discriminative Power}

%Many studies have utilized discriminative power to determine the stability of measures based on significance tests~\cite{dp1, dp2, dp3}, such as the paired bootstrap test\cite{dp}, Tukey's Honestly Significant Differences (HSD)\cite{HSD} test, and so on. Discriminative power can be used to estimate the necessary performance difference to establish statistical significance between two IR systems~\cite{DBLP:conf/www/Sakai12}. While discriminative power does not indicate whether a measure is correct or not, extremely low discriminative power does indicate that the metric is ineffective for drawing conclusions from an experiment~\cite{umeasure}.

\subsubsection{Intuitiveness}

The Concordance Test~\cite{DBLP:conf/www/Sakai12} is proposed to quantify the intuitiveness of diversity metrics. We believe it can also predict the intuitiveness of session-based metrics. 
We will first choose some golden standard measures and presume them to actually represent intuitiveness. 
Given a pair of metrics (M1, M2), the relative intuitiveness of M1(or M2) is computed in terms of preference agreement with the golden standard measures.

\subsection{Generating Runs for NST} \label{subsec:runs}

Since the organizers of NTCIR-16 Session Search (SS) Task have not released the run data of the participants, we use some ranking models to re-rank $NST$ to generate some runs for the experiments of intuitiveness. The models are comprised of: (1) \textbf{ad-hoc ranking models}, including KNRM~\cite{knrm}, ARC-II~\cite{arci}, Conv-KNRM~\cite{cknrm}, and DUET~\cite{duet}; 
(2) \textbf{session-based ranking models}, including HBA-Transformers~\cite{hba}, COCA~\cite{coca}, and RICR~\cite{ricr}. The settings of these models are all the same as in their original papers.

Moreover, since the current session-based ranking models are not advanced enough to consider the two assumptions we put forward, we artificially construct two types of runs based on these ranking models: 
%\begin{itemize}
(1) \textbf{Ideal runs.} For each query of the session, we first add the candidate documents of the subsequent queries in the same session into its pool. 
Then we re-rank the candidates using these models and keep the top ten results. 
We believe this can make the re-ranked session closer to the ideal session that we defined. 
%With these runs, we can test the intuitiveness of session-based metrics on the ``ideal sessions''.
(2) \textbf{Diversified runs.} For each query of the session, based on the extended candidate pool described above, we discard the candidate documents that are already presented (included in the top ten results) in the preceding queries. We believe this can make the session more ``diversified''.
%(as described in \textbf{A3}).

%\end{itemize}

We generate 7 original runs, 7 ideal runs, and 7 diversified runs based on 7 ranking models aforementioned for $NST$.

%print ('各个模型测试kendall tau:', np.mean(test_ks['sDCG/']), np.mean(test_ks['sRBP/']), np.mean(test_ks['U_Measure']),
%     np.mean(test_ks['sDCG/R']), np.mean(test_ks['sRBP/R']), np.mean(test_ks['sDCG/Q']), np.mean(test_ks['sRBP/Q']), np.mean(test_ks['nU_Measure']))
%各个模型测试spearman r: 0.03352698320844366 0.05084706052149418 -0.22821068748308096 0.34732759019816734 0.3507684366167333 -0.1800380032421676 0.31497823441615436 nan
%各个模型测试kendall tau: 0.024182218689926156 0.038247145334425185  -0.17003760656422698 0.2771369384849162 0.27948598543766967 0.24918993830444727 0.2506819638584701 nan2
% UM/Q : 0.1008124565771205, 0.07891635550270151
% NUM: 0.3594668561822826, 0.2873845066845209

\begin{table*}[tbp]
    \centering
    \caption{Spearman's $\rho$ and Kendall's $\tau$ between session-based metrics and user satisfaction on $FSD$. The best performance is in bold, and the second-best performance is underlined. "Improv." reflects the improvements of NUM over RS-RBP.}
    \begin{tabular}{p{0.13\textwidth}cccccccccc}
    \toprule
         {Metric} & {sDCG} & {sRBP} & 
         {sDCG/q} & {sRBP/q} & {U-measure} & {U-measure/q} & {RS-DCG} & {RS-RBP} & {NUM} & {Improv.} \\
        \midrule
        {Spearman’s $\rho$} & {{0.0335}} & {0.0508} & {0.3136} & {0.3150} & {-0.2282} & {0.1008} 
        & {0.3473} & \underline{0.3508} & \textbf{0.3611} & {2.94\%} \\
        
        {Kendall's $\tau$} & {{0.0242}} & {0.0382} & {0.2492} & {0.2507} & {-0.1800} & {0.0789} 
        & {0.2771} & \underline{0.2795} & \textbf{0.2884} & {3.18\%} \\
        
    \bottomrule
    \end{tabular}
    % }
    \label{tab:satisfaction}
\end{table*}

\begin{table}[t!]
    \centering
    \caption{Performance of ablated metrics on FSD.}
    \begin{tabular}{p{0.3\linewidth}ll}
    \toprule
        & Spearman’s $\rho$ & Kendall's $\tau$  \\
    \midrule
        {NUM w/o. SN} & -0.1960 \ \ -154.28\% & -0.1561 \ \ -154.13\%  \\
        {NUM w/o. RT} & 0.3276 \ \ -9.28\% & 0.2639 \ \ -8.50\%  \\
        {NUM w/o. SE} & 0.3522 \ \ -2.46\% & 0.2789 \ \ -3.29\%  \\
        %{NUM w/o. DE} & 0.3611 \ \ +0.44\% & 0.2884 \ \ +0.35\%  \\
        {NUM (Full)} & \textbf{0.3611} \ \ - & \textbf{0.2884} \ \ - \\
    \bottomrule
    \end{tabular}
    \label{tab:abla-satis}
\end{table}

\subsection{Baselines}\label{subsec:baseline}

To demonstrate the effectiveness of NUM and verify our two assumptions, we compare NUM with existing DCG-based and RBP-based metrics sDCG~\cite{sdcg}, sRBP~\cite{srbp}, sDCG/q, sRBP/q, RS-DCG~\cite{rsdcg} and {RS-RBP~\cite{rsdcg}}. 
In addition, to verify the effectiveness of the session-level normalization introduced in Section~\ref{subsubsec:normalize}, we also compare our metric with U-measure~\cite{umeasure} and U-measure/q.

Following the settings of~\cite{rsdcg}, supposing each query has $N$ documents and each session $S$ has $M$ queries, the computation of sDCG can be described as follows:
\begin{equation}
    {\rm sDCG}(S) = \sum_{m=1}^{M} \sum_{n=1}^{N} \frac{{\rm g}(d_{m,n})}{(1+{\rm log}_{b_q}m) (1+{\rm log}_{b_r}n)},
\end{equation}
where ${\rm g}(d_{m,n})$ maps the score of the $n$-th document in the $m$-th query of the session. And sRBP is computed as follows:
\begin{equation}
    {\rm sRBP}(S) = (1-p) \sum_{m=1}^{M} \left( \frac{p-bp}{1-bp} \right)^{m-1} \sum_{n=1}^{N} (bp)^{n-1} {\rm g}(d_{m,n}).
\end{equation}

``/q'' in sDCG/q and sRBP/q is a way of normalizing metrics by simply the number of queries in the session. 
%This is an alternative way to penalize query reformulations. 
It is described as:
${\rm Metric/q} = {{\rm Metric}(S)}/{M}$.

For the computations of RS-DCG and RS-RBP, we apply the settings of their original paper~\cite{rsdcg}:
\begin{equation}
    {\text {RS-DCG}}(S) = \sum_{m=1}^{M} e^{-\lambda(M-m)} \sum_{n=1}^{N} \frac{{\rm g}(d_{m,n})}{(1+{\rm log}_{b_q}m) (1+{\rm log}_{b_r}n)},
\end{equation}
\begin{equation}
    {\text{RS-RBP}(S)} = \sum_{m=1}^{M} e^{-\lambda(M-m)} \left( \frac{p-bp}{1-bp} \right)^{m-1} 
    \sum_{n=1}^{N} (bp)^{n-1} {\rm g}(d_{m,n}).
\end{equation}

Following~\cite{umeasure}, we calculate the $gain$ value of a $l$-relevant document as:
${\text g}(d) = (2^l-1)/{2^H}$,
where $H$ is the highest relevance level. 
Since this paper mainly discusses the scenarios where we only have implicit feedback, the $gain$ value of a session-level relevant document is $(2^1-1)/2^1=0.5$.
For NUM, the session-level labels are inferred from click-through data, according to A2.
For the other metrics, we treat the clicked document directly as relevant ones, and \textbf{do not} use A2 to enhance the corresponding relevance labels. 
%We will explore whether our assumptions hold in Section~\ref{subsec:abla}.

For U-measure, we build a trailtext from user clicks and compute U-measure on it as illustrated in Eq~(\ref{Eq:full-um}). 
The value of $L$ in~\cite{umeasure} is estimated using Microsoft's Bing (September 7, 2012, US market) data. 
However, we have to re-estimate $L$ due to the inconsistency between the language of Microsoft's Bing (English) and the datasets we utilize in this study (Chinese). 
Furthermore, we believe that we should take the search engine that the dataset uses into consideration when we estimate $L$.
%and the length of the reformulation text. 
This is because we believe that a user's tolerance for ``the largest amount of text that the user may have to read in one session'' can be different from one search engine to another. 
Thus we estimated $L$ independently for TianGong-SS-FSD~\cite{rsdcg} and NTCIR-16 Session Search (SS) Task~\cite{ntcir}. 
We estimated the MTL for each session by assuming that each snippet is 80 characters long (which is a reasonable assumption for Chinese search engines), and discarded 1\% of the sessions with the highest MTL values.
Note that we set $F$ as 20 following the original U-measure paper~\cite{umeasure}.

For the instantiations of the baseline metrics and NUM, we adopt different approaches with respect to different meta-evaluation techniques and the corresponding datasets. 
More details can be found in Section~\ref{subsec:satisfaction} and Section~\ref{subsec:intuitiveness}.
%\footnote{We will release the code based upon the acceptance of the paper.}
\textbf{Besides, we provide an anonymous version of our code for review.}\footnote{\url{https://anonymous.4open.science/r/NUM-C228/}}

\section{Results and Analysis}\label{sec:experiments}

\subsection{Correlation with User Satisfaction}\label{subsec:satisfaction}

We first compare the performance of NUM to the baseline metrics by computing Spearman's $\rho$~\cite{spearman} and Kendall's $\tau$~\cite{kendall} with user satisfaction on $FSD$.

For the instantiations of the baseline metrics and NUM in this experiment, we adopt a 5-fold cross-validation method following~\cite{rsdcg}. We repeat this approach ten times. 
For each time, we use one fold of data to test the metrics' correlation with user satisfaction and use the other four folds to tune the parameters. 
We tune the parameters of DCG-based and RBP-based baselines to fit user satisfaction based on Spearman's $\rho$. 
For DCG-based metrics, $b_r$ and $b_q$ are searched in range $(1.0,5.0]$ with step of 0.1. 
For RBP-based metrics, $b$ and $p$ are searched in range $(0,1)$.
The exact values of the parameters can be found in the anonymous version of our code.
For U-measure-based metrics (including NUM), we only estimate (not tune) their parameters ($L$ and the length of the reformulation text) because these parameters should be consistent with users' real reading behavior. 
Note that we estimate them on the same folds of data as other baselines for fair comparisons.

The average estimated $L$ is 19,336 for U-measure-based metrics.
Besides, We found that a query reformulation takes an average of 206 seconds, or 3.43 minutes, which is a relatively high cost.
Additionally, native Chinese speakers can usually read at an average speed of 255±29 words per minute~\cite{read_speed}. 
As a result, the average estimated length of the reformulation text is ($255\times3.43=875.5$) words (Chinese characters).

We report the average Spearman's $\rho$ and Kendall's $\tau$ between each metric and user satisfaction across all ten times of 5-fold cross-validation on $FSD$.
The results are shown in Table~\ref{tab:satisfaction}, which demonstrate the effectiveness of our method\footnote{We will notice that the results of some metrics are lower than those reported in~\cite{rsdcg}. This is because this paper mainly discusses real-world offline scenarios where we can only obtain implicit feedback, so we use the click-through data to estimate user satisfaction.}. Furthermore, we can make the following observations:

\textbf{(1) NUM achieves the best results among all metrics, demonstrating its effectiveness of estimating user satisfaction.}  
For example, when compared to the state-of-the-art baseline RS-RBP, our metric has improved Kendall's $\tau$ by around 3.18\%.
%We will notice that NUM's improvements over RsRBP are relatively small. We believe the reason is that the parameters of the baseline metrics are tuned using cross-validation, whereas NUM's parameters are estimated on the same data.

\textbf{(2) Compared to the original U-measure and the simply-normalized U-measure/q, NUM has a stronger correlation with user satisfaction, which demonstrates the effectiveness of the proposed session-level normalization.} We can observe that NUM outperforms the original U-measure, indicating that it is necessary to apply the session-level normalization and the click-through data enhancement for session-based evaluation.
Moreover, NUM performs better than U-measure/q, which further demonstrates the effectiveness of session-level normalization.

\subsection{Intuitiveness}\label{subsec:intuitiveness}

\begin{table}[t!]
    \centering
    \caption{Intuitiveness based on preference agreement with the proposed golden standard measures(AP and LCD). For each metric combination, the higher score is in bold, and the number of disagreements between these two metrics is stated in the parentheses below. The abbreviation "UM" stands for "U-measure".}
    \small
    \begin{tabular}{l|c|c|c|c|c}
    \toprule
    
         {\textbf{AP}}  & 
         {sRBP} & {UM} & {RS-DCG} & {RS-RBP} & {NUM} \\
        \midrule
        \multirow{2}{*}{sDCG} 
        & \textbf{0.89}/{0.83} & \textbf{0.89}/0.70 & \textbf{0.89}/0.65 & \textbf{0.84}/0.67 & 0.64/\textbf{0.91}  \\
        
        & (75,655) & (114,588) & (117,044) & (130,035) & (154,371) \\
        \midrule
        
        \multirow{2}{*}{sRBP} 
        & - & \textbf{0.85}/{0.71} & \textbf{0.86}/0.67 & \textbf{0.84}/0.69 & {0.62}/\textbf{0.92}  \\
        
        & - & (125,447) & (122,497) & (125,116) & (154,137) \\
        \midrule
        
        \multirow{2}{*}{UM} 
        & - & - & \textbf{0.86}/0.79 & \textbf{0.79}/0.78 & 0.58/\textbf{0.94}  \\
        
        & - & - & (96,944) & (115,498) & (172,019)\\
        \midrule
        
        \multirow{2}{*}{RS-DCG} 
        & - & - & - & {0.82}/\textbf{0.90} & 0.56/\textbf{0.94} \\
        
        & - & - & - & (72,863) & (182,192)\\
        \midrule
        
        \multirow{2}{*}{RS-RBP} 
        & - & - & - & - & 0.57/\textbf{0.92}  \\
        
        & - & - & - & - & (179,640)\\
    
    \midrule
        {\textbf{LCD}}  & 
         {sRBP} & {UM} & {RS-DCG} & {RS-RBP} & {NUM} \\
        \midrule
        \multirow{2}{*}{sDCG} 
        & 0.81/\textbf{0.86} & \textbf{0.93}/0.59 & \textbf{0.84}/0.62 & \textbf{0.75}/0.68 & 0.56/\textbf{0.85}  \\
        
        & (75,655) & (114,588) & (117,044) & (130,035) & (154,371) \\
        \midrule
        
        \multirow{2}{*}{sRBP} 
        & - & \textbf{0.92}/0.58 & \textbf{0.84}/0.61 & \textbf{0.77}/0.67 & {0.57}/\textbf{0.84}  \\
        
        & - & (125,447) & (122,497) & (125,116) & (154,137) \\
        \midrule
        
        \multirow{2}{*}{UM} 
        & - & - & {0.72}/\textbf{0.86} & 0.62/\textbf{0.88} 
        & 0.45/\textbf{0.94}  \\
        
        & - & - & (96,944) & (115,498) & (172,019)\\
        \midrule
        
        \multirow{2}{*}{RS-DCG} 
        & - & - & - & 0.73/\textbf{0.95} & 0.48/\textbf{0.87} \\
        
        & - & - & - & (72,863) & (182,192)\\
        \midrule
        
        \multirow{2}{*}{RS-RBP} 
        & - & - & - & - & 0.53/\textbf{0.83}  \\
        
        & - & - & - & - & (179,640)\\

    \bottomrule
    
    \end{tabular}
    % }
    \label{tab:intuitiveness}
\end{table}

Since there are few works trying to evaluate the intuitiveness of session-based metrics, the golden standard measures that represent the intuitiveness of session search have yet to be discovered. 
In this work, we suggest two metrics\footnote{For these metrics, we treat the clicked document directly as relevant ones and do not use A2 to enhance their labels. More details of these metrics can be found in our code.} for the intuitiveness test of session-level metrics: 

%(1) Session-level MAP \textbf{(S-MAP)}. We believe that MAP can represent intuitiveness for query-based evaluation because it attempts to make the search system rank the relevant documents higher than the irrelevant ones. 
%To apply MAP to session search, a straightforward way is computing MAP for each query of the session and averaging the results. 
%However, as described in {A1}, we believe that an ideal session should rank all the relevant documents before all irrelevant documents in the session.
%Thus we should evaluate the session as a whole rather than aggregate the results of queries. 
%As a result, we propose that we should treat the entire session as a virtual query and compute MAP on it as follows:
%\begin{equation}
%    {\text{S-MAP}}(S) = {\text{MAP}}(S).
%\end{equation}
%We hope that this can lead the search system to rank the relevant results higher than the irrelevant ones in the whole session, which we believe can represent the intuitiveness of the session search.
(1) Average Precision \textbf{(AP)}. 
We believe this measure is simple but intuitive, by which we simply compute the precision of each query and average these values.
%We believe that MAP can represent intuitiveness for query-based evaluation because it attempts to make the search system rank the relevant documents higher than the irrelevant ones. 
Note that because MAP depends on the recall base, which can not be estimated with click data, we thus use average precision instead. 
    
(2) The position of the Last Clicked Document \textbf{(LCD)}. This measure records the position of the last clicked document in the whole session and takes the reciprocal of this position as the score:
${\text{LCD}}(S) = {1}/{Index_{lc}}$,
where $Index_{lc}$ is the position of the last clicked document in the session $S$. 
This value depicts the number of snippets a user has to examine in order to obtain all the information she needs. 
For example, if each query has 10 candidate documents and the last clicked document of the session is the fourth document of the second query,  then $Index_{lc}=10+4=14, {\text{LCD}}(S)=1/14$.
We believe that the session with a higher LCD value should be preferred because the user can scan fewer snippets to complete her search task, saving her time and effort.
Thus, we believe LCD also represents the intuitiveness of session search.

\begin{table}[t!]
    \centering
    \caption{The ablation experiments of intuitiveness based on preference agreement with LCD. }
    %For each metric combination, the higher score is in bold, and the number of disagreements between these two metrics is stated in parentheses.}
    \begin{tabular}{p{0.3\linewidth}|p{0.25\linewidth}p{0.25\linewidth}}
    \toprule
       Metric & {NUM (Full)} & \# disagreements \\
        \midrule
        {NUM w/o. SN} 
        & 0.46/\textbf{0.93}
        & (157,785)  \\

        {NUM w/o. RT} 
        & 0.78/\textbf{0.99}
        & (89,495) \\

        {NUM w/o. SE} 
        & 0.75/\textbf{0.91}   
        & (96,245)\\

        %{NUM w/o. DE} 
        %& \textbf{0.92}/0.90 
        %& {(84,241)}\\
        
    \bottomrule
    
    \end{tabular}
    % }
    \label{tab:abla-intuitiveness}
    \vspace{-5px}
\end{table}
    
%(3) The Number of Queries used to find the last Clicked document \textbf{(NQC)}. 
%It counts the number of queries that the user has to issue to find the last clicked document and uses the reciprocal of this number to calculate the session's evaluation result. 
%For example, if the last clicked document of the session is the fifth document of the third query, then ${\text{NQC}}(S)=1/3$.
%We believe this number represents the number of queries the user must submit to complete her search task.

For the instantiations of the DCG-based and RBP-based metrics in this experiment, since there are no golden standard labels like user satisfaction in $NST$, we use the mean of the parameters tuned on each fold in Section~\ref{subsec:satisfaction} (their exact values can be found in the anonymous version of our code).
%we set their parameters as their original papers suggest~\cite{sdcg, srbp, rsdcg}. 
For U-measure-based metrics, we only need to estimate (not tune) $L$ as we explained in Section~\ref{subsec:satisfaction}. Thus, we use the remaining 145,154 sessions of $NST$ to estimate them and test the intuitiveness on the other 2,000 sessions. The estimated $L$ is 12,792.
For the length of the reformulation text, since there is no start and end timestamp of a query in $NST$, we simply use the estimated one in Section~\ref{subsec:satisfaction} (362).

We exclude sDCG/q, sRBP/q, and U-measure/q from this experiment because the ``/q'' normalization does not affect the concordance test (two runs have the same number of queries).
The concordance test is performed on all 21 runs (7 original runs + 7 ideal runs + 7 diversified runs, \ie, (21$\times$20/2$\times$2000= )420,000 session pairs. 

The results are presented in Table~\ref{tab:intuitiveness}. For example, the result at the top left represents that sDCG and sRBP disagree in 75,655 pairs of sessions.
Among these disagreed pairs, sDCG agrees with AP on around 89\% of them, while sRBP agrees on about 83\%, which implies sDCG is more intuitive than sRBP in terms of AP. Furthermore, we can observe that:

\textbf{(1) In terms of all golden standard measurements, NUM is more intuitive than all baselines.} For example, NUM agrees with AP on about 94\% of the 182,192 disagreement session pairs, whereas RS-DCG is only consistent with AP on around 56\%.

\textbf{(2) Incorporating the recency effect makes the metric less intuitive.} 
After incorporating the recency effect (RS-DCG and RS-RBP), we can observe that the intuitiveness of metrics decreases in terms of all golden standard measures. 
%For example, sRBP agrees with LCD on around 77\% of the 125,116 disagreement session pairs, whereas RS-RBP only agrees with LCD on about 67\% of them.
These findings reveal that, while the recency effect is beneficial for estimating user satisfaction, it degrades the metrics' intuitiveness.

\subsection{Ablation Study} \label{subsec:abla}

To further explore the reasonability of our two assumptions and the effectiveness of the improvements we make over the original U-measure, we design several variants of NUM.
Specifically, we conduct the ablation experiments on $FSD$ and $NST$ as follows:

%\begin{itemize}

$\bullet$  \textbf{NUM w/o. SN.} We remove the session-level normalization part (SN, illustrated in Section~\ref{subsubsec:normalize}). In another word, we only evaluate the actual session without considering the ideal session.

$\bullet$  \textbf{NUM w/o. RT.} We discard the reformulation text (RT, introduced in Section~\ref{subsubsec:rt}). We do not add extra empty texts into the trailtext to penalize query reformulations.

$\bullet$  \textbf{NUM w/o. SE.} We eliminate the click-through data enhancement, which states that the same document should be tagged relevant to the preceding query if it is clicked in a subsequent query but skipped in the previous one, \ie, the supplemental enhancement (SE) based on \textbf{A2}.

%\item \textbf{NUM w/o. DE.} We abandon the click-through data enhancement that the same document clicked in the preceding query and also clicked in the subsequent query should be marked irrelevant to the subsequent one, \ie, the diversified enhancement (DE) based on \textbf{A3}.

%\end{itemize}

The results of the ablation experiments are shown in Table~\ref{tab:abla-satis} and Table~\ref{tab:abla-intuitiveness}. 
%(Note that since the decreasing or increasing of discriminative power does not represent the effectiveness of the changes, we do not conduct ablation studies on discriminative power.) 
From which we can draw the following conclusions: 

\textbf{(1) Normalizing the evaluation result at the session level is effective.} 
In \textbf{A1}, we presume that in an ideal session, all relevant documents should be ranked before all irrelevant documents. 
%As a result, the user may not even need to reformulate her queries, saving her time and effort.
We can evaluate the similarity between this session and the ideal session by a session-level normalization. 
After removing this technique, our metric's intuitiveness and correlation with user satisfaction both drop. 
For example, Spearman's $\rho$ on FSD decreases by about 154.28\%. %and indicates a negative correlation if we remove the session-level normalization.
Furthermore, NUM without session-level normalization agrees with LCD on 46\% of the disagreement pairs, whereas NUM agrees with LCD on 93\%.
These declines show that normalizing the actual session's evaluation score with the ideal session's score is effective.
%in terms of both intuitiveness and estimating user satisfaction
It also supports \textbf{A1} that an ideal session should rank all relevant documents before all irrelevant documents in the session.
 
\textbf{(2) It is useful to include a reformulation text in the trailtext for each query reformulation.} 
We propose to add a reformulation text in the trailtext between every two queries (as illustrated in Section~\ref{subsubsec:rt}).
We believe it can penalize the query reformulation behavior to an adjustable degree, which the original U-measure does not take into account.
After eliminating the reformulation texts, the performance of our metric drops. 
For example, the performance of estimating user satisfaction decreases by about 8.50\% in terms of Kendall's $\tau$.
Furthermore, NUM agrees with LCD on about 78\% of the disagreement pairs without the reformulation text, whereas with the reformulation text, NUM agrees with around 99\%.
This indicates that penalizing query reformulations by adding empty texts into the trailtext is effective.

\textbf{(3) The supplemental enhancement (SE) makes our metric more intuitive and correlate better with user satisfaction.} 
To verify \textbf{A2}, we propose that if a document is clicked in a subsequent query but skipped in a preceding one, it should be tagged relevant to the preceding query.
After removing SE, the performance decreases.
For example, it causes a decrease of 3.29\% in terms of Kendall's $\tau$. 
Moreover, NUM agrees with LCD on about 75\% of disagreement pairs without SE, but the full NUM agrees with LCD on around 91\%.
These results demonstrate the effectiveness of SE and verify \textbf{A2}.
The reason these reductions are smaller than those of the preceding two removals is that the number of documents that require SE is small (approximately 1\% of $FSD$ and 3\% of $NST$).

%\textbf{(4) The diversified enhancement (DE) degrades the performance of NUM.} 
%To verify \textbf{A3}, we propose that if the same document has been clicked in a preceding query and is clicked again later, it should be marked irrelevant to the subsequent query.
%However, the performance actually increases after discarding DE.
%For example, it causes an increase of about 0.44\% in terms of Spearman's $\rho$.
%Besides, NUM agrees with LCD on about 92\% disagreement pairs without DE, whereas NUM (with DE) only agrees with LCD on around 90\%. 
%This suggests that \textbf{A3} is \textbf{not} helpful in terms of both estimating user satisfaction and intuitiveness. 
%A possible reason for \textbf{A3} being invalid is that users may hope to find their clicked documents remain in the top results during the search session, which can facilitate the re-finding behaviors and the trust of search systems as they can always rank relevant documents at top positions~\cite{nsdcg}.

\subsection{Dealing with Duplicate Documents} \label{subsec:duplicate}

As aforementioned in Section~\ref{subsec:click-enhancement}, there will be duplicate documents that are considered relevant in the ideal session after enhancing the click-through data based on A2.
There are three approaches to deal with the subsequent duplicate clicked documents:
(1) Exclude it from the trailtext; 
(2) Include it in the trailtext but give a discount to its relevance score; 
(3) Include it in the trailtext and do not give any discount.
For approach (2), 
%Yang et al.~\cite{DBLP:conf/ictir/YangL09}, proposed a discount factor that is tuned for different users.
since this discount factor is not our work's main contribution, we will simply set the discount value as 0.5 and leave more sophisticated discount factors to future work.

We implement these three approaches on $FSD$ and report the correlation with user satisfaction. 
The performances are presented in Table~\ref{tab:duplicate}.
We will notice that the difference between them is relatively trivial. 
This is because that the number of documents that require SE is small (approximately 1\% of $FSD$).
We can observe that approach (3) achieves best performance, which demonstrates that the duplicate documents produced by SE based on A2 should be include in the trailtext of the ideal session.
A possible reason is that the user may need to re-find information that she had already found or skimmed through earlier in the session.

\begin{table}[t!]
    \centering
    \caption{Performances of different approaches to dealing with duplicate documents on $FSD$.}
    \begin{tabular}{p{0.3\linewidth}|p{0.25\linewidth}p{0.25\linewidth}}
    \toprule
       Approach & Spearman’s $\rho$ & Kendall's \\
        \midrule
        {Exclude} & 0.3600  & 0.2876  \\

        {Include \& Discount} & 0.3597  & 0.2871  \\

        {No Discount} & \textbf{0.3611}  & \textbf{0.2884}  \\

        %{NUM w/o. DE} 
        %& \textbf{0.92}/0.90 
        %& {(84,241)}\\
        
    \bottomrule
    
    \end{tabular}
    % }
    \label{tab:duplicate}
    \vspace{-5px}
\end{table}

\section{Conclusions and Future Work}\label{sec:conclusion}

In this work, we identify two challenges in session-based evaluation and make two assumptions about evaluating a session. 
\textbf{A1} states that an ideal search system should rank all relevant documents before all irrelevant documents in the session.
\textbf{A2} believes that the documents clicked in a subsequent query but omitted in a preceding query are also relevant to that preceding query.
%\textbf{A3} states that the clicked documents of the previous queries are not relevant to the subsequent queries of the session due to redundancy. 
To verify these assumptions, we design a session-level metric called \textbf{Normalized U-Measure (NUM)}. 
NUM evaluates a session as a virtual query, uses the score of an ideal session to normalize the evaluation result (\textbf{A1}) and enhances the click-through data (\textbf{A2}). 
Experiments on two public datasets demonstrate that NUM is intuitive and able to estimate user satisfaction well. 
In addition, ablation studies demonstrate the effectiveness of \textbf{A1} and \textbf{A2}.

Our work may provide some guidance for designing better session-based evaluation metrics: 
(1) Future metrics may need to evaluate the session as a whole rather than aggregating each query's result. 
(2) Though we only incorporate the session-level normalization for U-measure in this study, this proposed technique can be applied to many session-based metrics. 
(3) It may be inappropriate to directly use user clicks as relevance judgment in real-world search scenarios. 
%The implicit feedback should be further enhanced to get session-level labels.
(4) Intuitiveness can be a good way to evaluate session-based metrics.

Nevertheless, our work still has some limitations that we aim to address in future work: 
(1) In this work, we only verify our assumptions through some experiments on two public datasets.
Some delicate user studies are needed to further verify the proposed assumptions and explore more assumptions about session-based evaluation. 
(2) We propose two assumptions and NUM under the condition of offline evaluation with implicit feedback. 
We will consider applying NUM to the scenarios where human relevance labels are available.  
(3) We propose one approach to infer session-level labels based on click-through data in real-world search scenarios.
There are more click-through data enhancement techniques to be discovered. 
For example, the click dwell time and eye-tracking information can be incorporated.
%into constructing better relevance labels.
(4) We propose two simple metrics that we believe can represent the intuitiveness of session search. 
It is only a primary step of exploring what metrics can serve as the golden standard measures for the intuitiveness of session search.

% \clearpage
% \bibliographystyle{ACM-Reference-Format}
% \bibliography{main}

\end{document}